\documentclass[11pt,thmsa]{article}
\usepackage{amssymb}
\usepackage{theorem}
\usepackage{amsmath}
\usepackage{epsfig}
\usepackage[]{algorithm}
\usepackage{algorithmic}
\usepackage[usenames,dvipsnames]{color}
  
\newcounter{theorem}
\theoremheaderfont{\sffamily\bfseries}
\theorembodyfont{\sffamily\slshape}
\newtheorem{theorem}{\sc Theorem}
\newtheorem{lemma}{\sc Lemma}

\newtheorem{definition}{\sc Definition}

\newtheorem{assumption}{\sc Assumption}

\newcommand{\Rea}{\mathbb{R}}
\newcommand{\Nat}{\mathbb{N}}

\newcommand{\Ex}[2]{\mathbb{E}_{#1}\left[#2\right]}

\usepackage{ulem}

\usepackage{natbib}
\begin{document}

\title{\vspace{-0cm}
{\bf Uniformly Self-Justified Equilibria}\thanks{
We thank Ken Judd, Filipe Martins-da-Rocha, Tom Sargent, Xinyang Wang, Bob Wilson, Thomas Winberry, as well as three anonymous referees, the editor, and seminar participants at various 
universities and conferences for useful conversations and comments.}
}
\author{
Felix Kubler \\
DBF, University of Zurich \\ 
and Swiss Finance Institute\\
fkubler@gmail.com
\and
Simon Scheidegger\\
Department of Economics, HEC Lausanne\\ 
and Enterprise for Society (E4S)\\
simon.scheidegger@unil.ch  
}

\date{\today}
\maketitle

\begin{abstract}
We consider dynamic stochastic economies with heterogeneous agents and introduce the concept of uniformly self-justified equilibria (USJE)---temporary equilibria for which forecasts are best uniform approximations to a selection of the equilibrium correspondence.
In a USJE, individuals' forecasting functions for the next period's endogenous variables are assumed to lie in a compact, finite-dimensional set of functions, and the forecasts constitute the best approximation within this set.
We show that USJE always exist and develop a simple algorithm to compute them.
Therefore, they are more tractable than rational expectations equilibria that do not always exist. As an application, we discuss a stochastic overlapping generations exchange economy and provide numerical examples to illustrate the concept of USJE and the computational method.

 \vspace{0.3cm}
 
 \vspace{0.1cm}

\end{abstract}

\clearpage

\section{Introduction}
\label{Sec:Intro}

Dynamic stochastic general equilibrium models with heterogeneous agents and incomplete financial markets play an important role in macroeconomics and public finance (see, e.g., \cite{krueger2016macroeconomics}). Unfortunately, for many model specifications, the computation of rational expectations equilibria is not possible.
A recursive equilibrium does not always exist (see \cite{kubler2002recursive}), and even when it does, no regularity properties (beyond measurability) can be established for the equilibrium policy functions (see \cite{brumm2017recursive}).  
Computational methods typically focus on computing rational expectation $\epsilon$-equilibria, that is, allocations and prices that clear markets and satisfy agents' optimality conditions (i.e., Euler equations) up to some \lq\lq small\rq\rq\ $ \epsilon\in \mathbb{R}_{++}$.
Errors in Euler equations provide a good method to analyze solutions to
dynamic optimization problems (see~\cite{santos2000accuracy} or~\cite{judd1992}). However, in models with heterogeneous agents, it is more difficult to interpret them. Agents' incorrect choices need to be coordinated to ensure that the definition of an approximate equilibrium holds. Agents' mistakes are determined by the requirement that at any time, agents' rational expectations from the previous period need to be correct in the current period, and that markets clear. Errors in the market clearing condition are similarly problematic.

These shortcomings seem to make a rigorous numerical analysis of rational expectations equilibria impossible and, in consequence, call for an alternative to the rational expectations assumption that is more tractable.\footnote{This refers to tractability in the sense of the first paragraph. We do not formally discuss computational tractability in this paper.} One way to solve this is to assume that agents are boundedly rational---as~\cite{sargent1993bounded} argues---a sensible approach to relax rational expectations is \lq\lq expelling rational agents from our model environment and replacing them with \textit{artificially intelligent} agents who behave like econometricians.\rq\rq \
In this paper, we develop an alternative to rational expectations equilibria and consider temporary equilibria (see, e.g.,~\cite{grandmont1977temporary}) with forecasting functions that are simple in the sense that they are easy to compute numerically.
These forecasting functions approximate (a selection of) the temporary equilibrium correspondence, but the agents might make significant mistakes in their forecasts. These stem from the definition of an equilibrium that we are proposing below:  The agents make no mistakes in choices, given their forecasts, and markets clear. Agents' forecasts are prevented from being arbitrarily accurate only because of the computational cost associated with more accurate forecasts.\footnote{We do not formalize the concept of \lq\lq computational costs\rq\rq \ in this paper. However, it is obvious that, for example, the use of a low degree polynomial for forecasting purposes comes at a lower computation cost than, for example, using a large number of radial basis functions.}

To deal with the aforementioned issues, this paper makes four contributions. First, we introduce a novel equilibrium concept---that of a \lq\lq uniformly self-justified equilibrium\rq\rq \ (USJE)---for stochastic general equilibrium models with heterogeneous agents. Second, we prove existence. Third, we develop an algorithm to approximate equilibria numerically. Fourth, we demonstrate the applicability of our new equilibrium concept and the corresponding numerical method in the context of a stochastic overlapping generations (OLG) model with many generations.

The basic idea of our proposed USJE approach is as follows: In a temporary equilibrium, agents use the current endogenous variables such as current prices and/or asset choices across all agents, and the exogenous shock, to forecast future marginal utilities for assets;  prices for commodities and assets in the current period ensure that markets clear. 
Forecasting functions are assumed to lie in a pre-specified, finite-dimensional class of simple functions (such as polynomials of a fixed degree), and an agent chooses a function that minimizes the maximum error of his forecast (i.e., that is the uniformly best approximation to equilibrium marginal utilities). In a USJE, the agents make optimal choices, given their forecasts, and the forecasts are (constrained) optimal, given equilibrium prices and choices.

We introduce the concept of USJE in an abstract framework that encompasses both economies with heterogeneous, infinitely lived agents, as well as economies with overlapping generations.
This general modeling framework allows us to investigate the properties of a USJE with as little notation as possible. We prove existence by relying on relatively weak assumptions on fundamentals and using well-known tools from mathematical economics.
In addition, we introduce a simple algorithm to compute a USJE numerically.
We then consider an exchange economy with overlapping generations to illustrate the abstract framework and to show that the assumptions necessary for existence translate to standard assumptions on fundamentals in a concrete economic model. Finally, we provide an explicit numerical example to illustrate the concept and our proposed algorithm in a simple framework.

In an earlier working paper---~\cite{kubler2019self}---we introduced a related concept of \lq\lq self-justified equilibria\rq\rq, where agents choose forecasts that minimize the average squared error along the equilibrium path. Unfortunately, this concept is theoretically much harder to analyze, and there are no existence proofs available. However, as we argue in that paper, it lends itself better to existing computational methods. The main reason for that is that least-squares approximation methods are standard in numerical analysis, and there is a large variety of methods that can be employed efficiently. Moreover, the least-squares approximation is not as delicate with respect to outliers as a uniform approximation.

There is a large and diverse body of work exploring deviations from rational expectation (see, e.g., \cite{RePEc:eee:jetheo:v:12:y:1976:i:3:p:455-471}, ~\cite{blume1982learning},~\cite{sargent1993bounded},~\cite{kurz1994structure},~\cite{woodford2013macroeconomic},~\cite{gabaix2014sparsity},~\cite{adam2016stock}).
Much of this literature is motivated by insights from behavioral economics about agents' behavior or the search for simple economic mechanisms that enrich standard models' observable implications. Much of this research strand also focuses on single-agent models, where many of the technical difficulties discussed in this paper disappear.
Our work's motivation is very different in that we develop a simple alternative to rational expectations that allows researchers to rigorously analyze stochastic dynamic models with heterogeneous agents. The existence of an equilibrium and the tractability of the equilibrium functions seem to be a prerequisite for this.
Our concept of a  USJE  is somewhat similar to the idea of Berk-Nash equilibria discussed in~\cite{esponda2016berk}. However, the focus of their paper and much of the literature that follows is on learning, whereas ours is on the existence and simplicity of the equilibrium functions. Moreover, to the best of our knowledge, there is no other concept in the literature that ensures the general existence and tractability of an equilibrium for models with incomplete markets and heterogeneous agents.

The remainder of the paper is organized as follows. In Section~\ref{sec:model}, the general economy is introduced. Furthermore, we define uniformly self-justified equilibria and show that they exist under general conditions. Finally, we develop a simple algorithm to approximate USJE numerically. In Section~\ref{secolg}, we demonstrate how the concept of USJE and the related algorithm can be applied in practical applications. To this end, we consider an exchange economy with overlapping generations. Section~\ref{seccon} concludes.

\section{A general dynamic Markovian economy}
\label{sec:model}

Time is discrete and indexed by $t \in \mathbb{N}_0$. Exogenous shocks $ z_t $ realize in a finite set  $ {\mathbf Z} =\{1, \ldots, Z \}$, and follow a first-order Markov process with transition probability $ (\pi(z'|z))_{z,z' \in {\mathbf Z}} $.
A history of shocks up to some date $ t $ is denoted by $ z^t = (z_0,z_1,\ldots,z_t)$ and called a \lq\lq date event\rq\rq.  For $ \tau > t $, we write $ z^{\tau} \succ z^t $ if a date event $ z^{\tau} $ is a successor of  $ z^t $.

\subsection{An abstract economy}

The following abstract setup is borrowed from \cite{duffie1994stationary} and \cite{kubler2003stationary}.
It is useful to assume that each agent that is active in markets in a given period makes a current decision, given expectations over the next period's endogenous and exogenous variables such as the next period's choices and prices. These expectations are formed based on the current
endogenous variables and the exogenous shock. While these expectations are always correct in a rational expectations equilibrium, we allow them here to be imprecise (in a way which we will formalize below).
The endogenous variables at a given date event, $
\left(x(z^t) ,  y (z^t) \right) \in {\mathbf X} \times {\mathbf Y},  $ consist of variables, $ x(z^t) \in {\mathbf X} $, that are predetermined (possibly stochastically) by the previous period's choices,  as well as the current period's choices and prices $ y(z^t)  \in {\mathbf Y} $,
where in our setting both $ {\mathbf X} $ and $ {\mathbf Y} $ are subsets of the Euclidean space.
As in \cite{duffie1994stationary}, the expectations correspondence, $$ {\mathbf H} : {\mathbf X} \times {\mathbf Y}\times {\mathbf Z} \rightrightarrows ({\mathbf X} \times {\mathbf Y})^Z, $$ embodies all  short-run equilibrium conditions, that is, inter- and intra-temporal first order conditions as well as market clearing conditions. If agents expect the next period's prices and choices to lie in this correspondence, they have no incentives to deviate from the current choices.
We focus on environments where these conditions are sufficient for optimality, that is, if at all nodes $ z^t $,
\begin{equation*}
(x(z^{t+1}), y(z^{t+1}))_{z_{t+1} \in \mathbf Z}
 \in {\mathbf H} (x(z^t), y(z^t),z_t),
\end{equation*}
then the process $ (x(z^t),y(z^t))_{z^t} $ constitutes a competitive equilibrium.
For a given economy, there can be several different ways to write the expectations and to fix the set of endogenous variables, $ {\mathbf X} $, for example, one can include continuation utilities. This turns out to make a difference for our equilibrium concept defined below.
 
We define $ s=(x,z) \in {\mathbf S}={\mathbf X} \times {\mathbf Z} $ to be the state of the economy, and refer to $ x \in {\mathbf X} $ as the endogenous state, and to $z\in {\mathbf Z}$ as the exogenous shock. At $ t=0 $, the state $ s_0 $ is given, and we refer to it as the initial conditions of the economy.
A transition function $T$ maps the state today and actions in the current period to a probability distribution over states in the subsequent period: 
\begin{equation*}
    T: {\mathbf X} \times {\mathbf Y} \times {\mathbf Z} \rightarrow {\mathbf X}^Z.
\end{equation*}
The latter is exogenously given and describes how choices in the current period lead to predetermined variables in the next period. Note that in Section~\ref{secolg} below, we will provide concrete examples for the expectations correspondence and the transition function.

A (forecasted) policy function\footnote{In our framework, all policy functions represent the agents' expectations about future optimal choices and prices. We sometimes refer to these as \lq\lq forecasts\rq\rq, or \lq\lq forecasted policies\rq\rq.} maps states to current endogenous variables, that is,  
$$ P : {\mathbf X} \times {\mathbf Z} \rightarrow {\mathbf Y} .$$ 
This function describes an agent's expectations about which equilibrium prices and allocations arise from the predetermined variables and the shock. The temporary equilibrium correspondence, ${\mathbf N}_P$, maps the exogenous shock and the predetermined variables to  all endogenous variables that are consistent with an equilibrium (more precisely, with the expectations correspondence) in the current period, given that the next period's choices and prices are expected to be described by the policy, $P$:
 $$ {\mathbf N}_P: {\mathbf S} \rightrightarrows {\mathbf Y} , $$ with
$$ {\mathbf N}_P(x,z)=\{y \in {\mathbf Y}: \left( T_{z'}(x,y,z),P(T_{z'}(x,y,z),z') \right)_{z' \in {\mathbf Z}}  \in {\mathbf H}(x,y,z) \}. $$
In this recursive formulation, we drop the time-index, $t$, and superscript variables in the subsequent period with primes.
While in our environment, the transition function is exogenously given and therefore known, the policy function embodies the agents' expectations about the future. The policy function is defined as part of our equilibrium concept.
In a recursive (rational expectations) equilibrium, $ P(x,z) \in {\mathbf N}_P(x,z) $ for all $ (x,z) $. \
In contrast, in a USJE, this strong requirement is relaxed, and 
agents approximate the equilibrium correspondence by a \lq\lq simple\rq\rq\ function which we will assume to be the weighted sum of a given number of basis functions. In our examples below, we use polynomial basis functions, but there is a large literature in approximation theory on basis function expansions (\cite{friedman2001elements}) which explores many alternatives.
 We take $ {\mathbf Y} \subset \Rea^Y $, and assume that each $ P_k(\cdot) $, $ k=1,\ldots,Y $, is a \lq\lq best uniform approximation\rq\rq \ of (a selection of) the equilibrium correspondence within our given class of simple functions.\footnote{In general, we cannot ensure the uniqueness of the best uniform approximation. Therefore we refer to our approximations as {\it a best}.}
For each $ k=1,\ldots, Y $, the class of these simple functions is denoted by $ {\mathbf P}_{k} $ and is exogenously given. Our interpretation of this assumption is that economic agents cannot evaluate arbitrarily complicated functions and resort to approximations to decrease computational costs.
We define a USJE as follows:
\begin{definition}
\label{USJE}
	A uniformly self-justified equilibrium is given by a policy function, $$ P^*:{\mathbf X} \times {\mathbf Z} \rightarrow {\mathbf Y}, $$
	a finite set of points $ (x_z^1,y_z^1),\ldots, (x_z^N,y_z^N) $,
	for each $ z \in {\mathbf Z} $, with $ x_z^i \in {\mathbf X} $, $ y_z^i \in {\mathbf N}_{P^*}(x_z^i,z) $, for all $ i=1,\ldots, N $, and a set of states $ \widehat{\mathbf S} \subset {\mathbf S} $
	that satisfy the following two properties:
\begin{enumerate} 
	\item For each $ z \in {\mathbf Z} $ and all $ k=1,\ldots, Y$, the policy function, $ P^*_k(\cdot,z) $, is  a best uniform approximation to the points $ (y_{kz}^i)_{i=1,\ldots,N}$, that is,
	$$ P^*_{k}(\cdot,z) \in \arg \min_{f \in {\mathbf P}_k} \sup_{i=1,\ldots,N} | y^i_{zk}-f(x^i,z) | .$$ 
	\item 
	For each $ s=(x,z)\in \widehat{\mathbf S} $, there is a $ y(s) \in {\mathbf N}_{P^*}(s) $ such that for all $ k=1,\ldots,Y$, 
	$$ | P^*_k(s)-y_k(s) | \le  \sup_{i=1,\ldots,N} | y^i_k-P^*_k (x^i_z,z) |, $$
	and
	$$ (T_{z'}(x,y(s),z),z') \in \widehat{\mathbf S} \mbox{ for all } z' \in {\mathbf Z}. $$
\end{enumerate}
\end{definition}

In this definition~\ref{USJE}, contrary to what one might expect, the finite set of points $ (x^i_z,y^i_z) $  do not need to satisfy $ (x^i_z,z) \in \widehat{\mathbf S} $. Depending on how the set $ \widehat{\mathbf S} $ relates to $ {\mathbf S}$, this fact can lead to equilibria where for the realized equilibrium variables, the forecasted policy, $ P(\cdot) $, is not a best uniform approximation. However, as we argue below, it is computationally not feasible to require that $ x^i \in \widehat{\mathbf S} $. Of interest is then  the distance of the points  $ x^i $ to equilibrium states, that is,
$$ \max_{i=1,\ldots,N}\inf_{s \in \widehat{\mathbf S}} \| x-x^i \|
.$$
In our computational method below, we require that for some small $ \eta > 0 $ and for each $i$  and $z$, there is a $ s=(x,z) \in {\mathbf S} $ such that 
 $$  \| x- x^i_z\| \le \eta . $$

As first pointed out by \cite{hellwig1983note} (see also \cite{kubler2002recursive}), recursive equilibria do not always exist. The fact that typically the temporary equilibrium correspondence cannot be shown to be single-valued implies that one might not be able to always find a selection that is consistent with the last period's expectations.
While recursive equilibria may fail to exist, a USJE exists under mild (and standard) assumptions on fundamentals (similar to the assumptions in \cite{kubler2003stationary}). We give a general existence proof here and, in Section \ref{secolg}, provide detailed
assumptions on economic fundamentals that guarantee existence in the concrete case of a stochastic OLG model.

\subsection{Assumptions}
\label{Sec:Assumptions}

The following assumption describes the set of admissible approximations to the next period's policy. In the framework of a USJE, these are  fundamentals of the economy:

\begin{assumption}
\label{ass1}
For each $ k=1,\ldots, Y$, there are $ D $ continuous basis functions 
$$ \psi_{dk}: {\mathbf X}   \rightarrow \Rea , d=1,\ldots,D ,$$ 
such that
$$  {\mathbf P}_k =\{f : \  f(\cdot) = \sum_{d=1}^{D} \alpha_{d} \psi_{dk}(\cdot), \quad
\alpha \in \Rea^{D}   \}. $$
\end{assumption}
In assumption~\ref{ass1}, the number of basis functions is assumed to be the same for each $ {\mathbf P}_k $, $k=1,\ldots,Y $. It is conceptually easy to allow for different approximations for different policies, that is to say, to have $D$ differ across policies $k=1,\ldots,Y $. This extension might be interesting for applications where part of the agents' heterogeneity stems from their ability to forecast. However, since the notation becomes cumbersome, we omit this extension here.  Nevertheless, observe that the basis functions depend on $k$ and can vary with $k$.

Note that, depending on the approximation, it is not guaranteed that every $f \in {\mathbf P} $ maps into the set $ {\mathbf Y} $ as required by the definition of a policy function (cf. definition~\ref{USJE}). As we will see below, this turns out not to be problematic in practice.  For now, we can think of the expectations correspondence projecting $ P(x,z) $ into $ {\mathbf Y} $ when necessary. We will give an example for this construction in Section~\ref{secolg} below.

Assumption~\ref{ass1} allows us to write the temporary equilibrium correspondence, ${\mathbf N}_P(\cdot) $, as a map from $(x,z) \in {\mathbf S} $ and $ \vec \alpha \in \Rea^{DYZ} $, where  each $ \vec \alpha_{k,z} $ denotes the vector of $D$ coefficients of the policy $ P_k(\cdot,z) $. We then omit the subscript $P$, since the policy is determined by the coefficients $ \vec \alpha $.
As we will show in our OLG example in Section~\ref{secolg} below, the equilibrium correspondence will be upper-hemi-continuous and non-empty value in these coefficients. Assumption~\ref{ass2}  states this requirement for our abstract setting:

\begin{assumption}
	\label{ass2}
	\mbox{}
	\begin{enumerate}
		\item Sets $ {\mathbf X} $  and $ {\mathbf Y} $ are compact and, the equilibrium correspondence ${\mathbf N} (.) $  has a closed graph as a correspondence in $ s \in {\mathbf S} $ and in coefficients $ \vec \alpha \in \Rea^{DYZ} $. 
		\item There is a compact $ {\mathbf A} \subset \Rea^{DYZ} $ such that, 
			given any finite  $\widehat{\mathbf X} \subset {\mathbf X} $,  
			there exist policies, $ P_k=P_k(\cdot,\vec \alpha)  $, $ k =1,\ldots Y$,  and $ y(x,z)\in {\mathbf N}_P(x,z) $ for each $ x \in \widehat{\mathbf X} $	and each $ z \in {\mathbf Z} $ such that
			$$ P_k(.,z) \in \arg\min_{f \in {\mathbf P}_k} \sup_{x \in \widehat{\mathbf X}} \| y_k(x,z)-f(x,z) \| , \quad k=1,\ldots, Y, $$ and
			such that $ \vec \alpha \in {\mathbf A} $.
	\end{enumerate}
	
\end{assumption}

This assumption is very abstract---in the OLG model below (cf. Section~\ref{secolg}), we show that it can be derived from fundamentals.

\subsection{Existence}
\label{sec:existence}

The main result of this section reads as follows:
\begin{theorem}
\label{thm1}
	Under Assumptions \ref{ass1} and \ref{ass2}, there exists a uniformly self-justified equilibrium.
\end{theorem}

The proof consists of two parts. 
As a first step, we find a convenient finite-dimensional representation of self-justified equilibria.
In the second step of the proof, we consider a sequence of points that becomes dense in the true state space and show that (for a convergent subsequence) the associated equilibria converge to a USJE.

\subsubsection{Best uniform approximations}

The key to establishing existence lies in the following simple facts about the best uniform approximation. Let $ \{(x^1,y^1),\ldots,(x^N,y^N) \} $ be a finite set of $N$ points in $ \Rea^{n}\times \Rea $. 
Let $ \psi:\Rea^n \rightarrow \Rea^D $, where the $ \psi_d(\cdot)$, $ d=1,\ldots, D$,  denote basis functions as in Assumption \ref{ass1} above.
For $ N>D $, consider the problem:
\begin{equation}
	\label{fiua}
	\min_{\alpha \in \Rea^D} \sup_{i=1,\ldots,N} |y^i-\sum_{d=1}^D \alpha_d \psi_d(x^i) |.
\end{equation}
As~\cite{sukhorukova2018chebyshev} point out, expression~\eqref{fiua}
is a convex optimization problem, and its solution can be characterized by the requirement that zero lies in the subgradient of a function $ g(\cdot) $ that maps coefficients $ \alpha \in \Rea^D $ to positive numbers, that is,
$$	g(\alpha)= \sup_{i=1,\ldots,N} |y^i-\sum_{d=1}^D \alpha_d \psi_d(x^i)  | ,  $$
at the optimal solution of the minimization problem. By Caratheodory's theorem (\cite{sukhorukova2018chebyshev}),  there must be
$ D+1 $ points in the subgradient that span zero, that is to say, there exist $ D+1 $ points $ (\hat x^1, \hat y^1),\ldots, (\hat x^{D+1}, \hat y^{D+1}) \in \{(x^1,y^1),\ldots,(x^N,y^N) \}$ such that
for all $ j=1,\ldots, D+1 $, 
\begin{equation}
	\sup_{i=1,\ldots, N} |y^i-\sum_d \alpha_d \psi_d(x^i) | =|\hat y^j-\sum_d \alpha_d \psi_d(\hat x^j) |,
\end{equation}
and such that 
\begin{equation}
	0 \in \mbox{ convex hull} \cup_{j=1}^{D+1} \{  \mbox{sgn}(\hat y^j-\sum_d \alpha_d \psi_d(\hat x^j)) \cdot \psi(\hat x^j) \}.
\end{equation}
Note that this condition is necessary and sufficient for
$\sum_d \alpha_d \psi_d(\cdot) $ to be the best uniform approximation for $ (x^i,y^i)_{i=1}^N $.
This is true even if $ x^i=x^j $ for some $ i,j $ with different associated values $ y^i$, and $y^j $. In fact, in this case, zero might be the combination of only two points in the subgradient, namely if $ y^i $ and $ y^j $ are extreme points on opposite sides of the approximating function.

Therefore, in the context of our abstract economic setting, we can state the following lemma:
\begin{lemma}
	\label{crulem}
		Given any finite  $ \widehat{\mathbf X} \subset {\mathbf X} $,  the following two statements are equivalent:
	\begin{enumerate}
		\item
	There are policies, $ P  $, and a selection of the equilibrium correspondence, $ y(x,z)\in {\mathbf N}_P(x,z) $ for each $ x \in \widehat{\mathbf X} $	and each $ z \in {\mathbf Z} $ such that for each $ k=1,\ldots,Y $,
	$$ P_k(.,z) \in \arg\min_{f \in {\mathbf P_k}} \sup_{x \in \widehat {\mathbf X}} \| y_k(x,z)-f(x,z) \| $$
		\item For
	each $ z \in {\mathbf Z} $ and each $ k=1,\ldots, Y$, there are finite sets $ {\mathbf X}^F_{kz} $ consisting of
	$ D+1 $ points $ \hat x \in  \widehat{\mathbf X} $,
	such that
	$$ \sup_{x \in \widehat{\mathbf X}} | P_k(x,z) - y_k(x,z) | = 
	| P_k(\hat x,z)-y_k(\hat x,z) | ,\mbox{ for all } \hat x \in {\mathbf X}^F_{kz} $$
	and
	$$ 0 \in \mbox{ convex hull } \cup_{\hat x \in {\mathbf X}^F_{kz}} \{ 
\mbox{sgn}	(P_k(\hat x,z)-y_k(\hat x,z)) \cdot
	\psi(\hat x) \}. $$
	\end{enumerate}
\end{lemma}

This result is remarkable for two reasons. First, the characterization of the equilibrium is independent of the number of points in the finite set $\widehat {\mathbf X}$, and second, because the condition is necessary and sufficient.

\subsubsection{Proof of the Theorem 1}
\label{sec:proof_of_theorem}

We consider a nested sequence of finite sets of points $ (\widehat{\mathbf X}_i)_{i=1}^{\infty} $, where
$  \widehat{\mathbf X}_i $ becomes dense in $ {\mathbf X} $ as $i \rightarrow \infty$.
As per Assumption~\ref{ass2} part 2, for every $i$ and $ \widehat{\mathbf X}_i $, there exists a $ P^{i} $ and $ y(x,z) \in {\mathbf N}_{P^i}(x,z) $ for every $ x \in \widehat{\mathbf X}_i $, $ z \in {\mathbf Z} $ that satisfies
for each $ k=1,\ldots, Y,$
$$ P^i_k(\cdot,z) \in \arg\min_{f \in {\mathbf P}_k} \sup_{x \in \widehat {\mathbf X}_i} | y^i_k(x,z)-f(x,z) | .$$ 
By Lemma~\ref{crulem}, each $ P^i_{kz} $, $ k=1,\ldots, Y, $ $ z \in {\mathbf Z}$, can be characterized by
$  (D+1) $ points, $ 
x^{*i}_{k,z,l} \in \widehat{\mathbf X}_i $, $ l=1,\ldots,D+1$,  coefficients $ \alpha_{k,z} \in {\mathbf A}$ as well as a selection of
the equilibrium correspondence $ y^{*i}(x^{*i}_{k,z,l},z) \in {\mathbf Y} $. 
Assumption \ref{ass2} guarantees that $ {\mathbf X}, {\mathbf Y} $, and $ {\mathbf A} $ are all compact sets. Therefore, as $ i \rightarrow \infty $, there must be a convergent sub-sequence converging to
some $ (\alpha^*_{k,z})_{ k=1,\ldots, Y, z \in {\mathbf Z}} $, and some $ (x^{*}_{k,z,l})_{ k=1,\ldots, Y, z \in {\mathbf Z}, l=1,\ldots, D+1} $.
The coefficients, $ \alpha^*$, define forecasting functions $ P^*$.
By upper-hemi continuity (Assumption \ref{ass2}, part 1), for all $ z \in {\mathbf Z}, \ k=1,\ldots, Y $, if for every $i$, 
$$ 0 \in \mbox{ convex hull } \cup_{\hat x \in \{ x_{k,z,l}^{*i}, l=1,\ldots, D+1 \} } \{ 
\mbox{sgn}(P^{i}_k(\hat x,z)-y_k(\hat x,z)) \cdot
\psi(\hat x) \}, $$
then this must also hold in the limit, that is,
$$ 0 \in \mbox{ convex hull } \cup_{\hat x \in \{ x_{k,z,l}^{*}, l=1,\ldots, D+1 \} } \{ 
\mbox{sgn}((P^*_k(\hat x,z)-y^*_k(\hat x,z)) \cdot
\psi(\hat x) \}, $$ 
for some $ y^*(\hat{x},z) \in {\mathbf N}_{P^*}(\hat x,z) $.
Therefore, $ P^* $ satisfies part 1 of the definition of a USJE.
For any (other) $ x\in {\mathbf X} $, since $ \widehat{\mathbf X}_i $ is becoming dense in $ {\mathbf X} $, there must be a sequence $ x^i \rightarrow x $ with some $ y^i \in {\mathbf N}_{P^i}(x^i,z) $ that satisfies
$$ |y_k^i - P_k^i(x^i,z) | \le  \sup_{x \in \widehat{\mathbf X}_i} | y_k^i(x,z)-P_k^i(x,z) | .$$
By upper-hemi continuity, there must be a $ y \in {\mathbf N}_{P^*}(x,z) $ such
that
$$ |y_k^i - P_k(x^i,z) | \le  \sup_{l=1,\ldots, D+1} | y^i_k(x^*_{k,z,l},z)-P^*(x^*_{k,z,l},z) | .$$
This completes the proof of Theorem~\ref{thm1}.

\subsection{A computational method to approximate an USJE}
\label{sec:comp-USJE}

We now introduce two algorithms that will demonstrate how to compute a UJSE numerically. The first algorithm computes a solution to a discretized problem (whose existence was assumed in Assumption \ref{ass2} part 2). The second algorithm finds extreme points in the sense of the first part of the definition of a USJE.

To approximate a USJE numerically, an important obstacle seems to lie in the fact that finding a uniform approximation to a continuous function
is a difficult task. While in the univariate case, the so-called \lq\lq Remez algorithm\rq\rq \  
\citep{remes1934procede,pachon2009barycentric} 
provides an efficient tool for computing the best approximation, matters become complicated in higher dimensions. As~\cite{reemtsen1990modifications} points out, one of the main challenges, in this case, is to find global optima of the error function.

In our setting, the definition of a USJE implies that this is not necessary for obtaining a best uniform approximation since we do not require $P$ to be a best approximation on $ {\mathbf S}$. We merely need to find points in $ {\mathbf S} $ together with endogenous variables, for which $P(\cdot)$ is a best uniform approximation, and we need to ensure that along any simulated path, the maximum approximation error remains smaller than the error from this approximation.

A crucial step of the method is to obtain a best uniform approximation for the discretized problem (see, e.g.,~\cite{reemtsen1990modifications}).
As~\cite{stiefel1960note} points out, a best uniform approximation to a finite number of points can be obtained by solving a simple linear programming problem. 
The problem~\ref{fiua} can be written as 
\begin{eqnarray}
\label{LP}
&& \min_{\alpha \in \Rea^D, w \ge 0} w \quad \mbox{s.t.}\\ \notag
	&& y_i-\sum_{d=1}^D \alpha_d \psi_d(x^i)  \le w \ i=1,\ldots, N \\ \notag
	&& - y_i+ \sum_{d=1}^D \alpha_d \psi_d(x^i)  \le w \ i=1,\ldots, N \quad ,
\end{eqnarray}
which can be solved efficiently with standard methods~\citep{schrijver1998theory}, even when the number of points, $N$, becomes large.
				\begin{algorithm}[t!]
			\caption{\textit{Time iteration for the discretized problem}}
			\label{alg2}\vspace{0.0cm}
			\begin{enumerate}
				\setcounter{enumi}{-1}
				\item The input consists of a set of states $ (s_t)_{t=1}^n $, current policy function $P:{\mathbf S} \rightarrow {\mathbf Y}$, and tolerance $\bar{\epsilon} > 0$.
				\item Solve the equilibrium conditions at $n$ points:
				For all $t=0,\ldots,n$,
				solve the equilibrium conditions at  $ s_t=(x,z) $, 
			 \begin{equation}
			      \label {ecs}
			 \left( T_{z'}(x,y,z),P(T_{z'}(x,y,z),z') \right)_{z' \in {\mathbf Z}}  \in {\mathbf H}(x,y,z),
			  \end{equation}
           for $y \in {\mathbf Y}$, and set $ \hat x (z^t)=x $, $\hat y(z^t)=y $.
          		\item Update the policy function:  \\   	
				Define new policies $\tilde{P}_k(\cdot,z)$, $ k=1,\ldots Y$,
				$ z \in {\mathbf Z} $ by solving the linear program (\ref{LP}) for $\left\{\hat x(z^t) , \hat y_k(z^t)\right\}_{z_t= z}$.
				\item Check stopping criterion: \\
				Compute the error: $\epsilon = \| \tilde{P}-P \|_\infty / \| P \|_\infty$. Set $P=\tilde{P}$. If $\epsilon < \bar{\epsilon}$, 
				stop and report $P$; else go to step 1.
			\end{enumerate}
		\end{algorithm}
%
Algorithm~\ref{alg2} solves for a best uniform approximation of the policy in a USJE that is defined on a finite set of points. It corresponds to a standard time iteration algorithm as used, for example, in collocation methods (see, for example,~\cite{Krueger20041411}).

As the next step, we use in Algorithm~\ref{alg3} the ideas from the exchange algorithms of~\cite{remes1934procede} and~\cite{reemtsen1990modifications} and, along several simulated paths of fixed length, determine points where the maximum forecasting error exceeds the previously computed maximum. If this is the case, the point that yields the lowest forecasting error is exchanged with the new point.
				\begin{algorithm}[htb]
			\caption{\textit{ USJE exchange algorithm}}
			\label{alg3}\vspace{0.0cm}
			\begin{enumerate}
				\setcounter{enumi}{-1}
				\item Draw initial points, $ (s^t)_{t=0,\ldots,n} $, $ s^t \in {\mathbf S} $ for all $ t$,
				and use Algorithm \ref{alg2} to obtain initial $P(\cdot) $.
				For all $ k=1,\ldots Y$, $ z \in {\mathbf Z} $,
				$ t=0,\ldots, n$, define initial $ \hat x^t_{kz}=x^t. $
				Initialize $ \mbox{maxerr}_{kz} = 0 $ and
				$ \mbox{err}_{kz}^t=0 $ for all $t$.
				\item Draw a sequence of $ m \gg n $ random shocks $ (z_t)_{t=1}^m $.
				\item Solve equilibrium conditions along a simulated path. For all $t=0,\ldots,m$ 
				\begin{enumerate}
            \item Solve equilibrium equations (\ref{ecs}) at  $ x=x(z^t),z=z_t $ for $y$  and set $ x(z^{t+1}) = T_{z_{t+1}}(x,y,z) $.
				\item Update the points:  \\   	
				For all $ k=1,\ldots,Y $, 
				if $  |y_k-P_k(x,z)| \ge \mbox{maxerr}_{kz} $ 
				search for a point in $ \tilde{x} $ with $ \| x-\tilde{x} \| \le \eta $ and a solution of the equilibrium equations (\ref{ecs}), $ \tilde y$, that satisfy
				$$ |\tilde{y}_k-P_k(\tilde{x},z)| > |y_k-P_k(x,z)| .$$ 
				Search for $i$ that solves
				$$ \min_{i=1,\ldots,n} \mbox{err}_{kz}^i ,$$
				replace $ \hat{x}^i_{kz}  $ by $ \tilde{x} $ and 
				set 
				$$ \mbox{err}_{kz}^i = |\tilde{y}_k-P_k(\tilde{x},z)|.  $$ 
\end{enumerate}
				\item Find new policy by applying Algorithm \ref{alg2} on new points  $ \hat x^t_{kz} $, $ k=1,\ldots,Y $, $ z \in {\mathbf Z}$, $ t=0,\ldots,n$. 
				
				\item Check stopping criterion: \\
				If no points where exchanged in step 2.(b)  
				stop and report $P$; else go to step 1.
			\end{enumerate}
		\end{algorithm}
%
Step 2.(b) of the algorithm contains the basic idea. If, along a simulated path, the previous maximum forecasting error is exceeded, we find a new point with a new maximum error. In an idealized setup, this point would maximize the error within some small neighborhood of $x$, but this is obviously computationally quite expensive. It is not important to find a (local) maximum, but rather to find a point where the error is larger than at $x$. In the practical application below, we find this point by taking one gradient ascent step of pre-described length with possible back-tracking. 
		
While a USJE always exists, there is no general proof that Algorithm~\ref{alg3} converges. However, in our numerical examples below, this is always the case. Note that in the computational method, we neglect the problems that arise with the possibility of multiple solutions to expression~\eqref{ecs}.
		
We propose a simple error analysis to evaluate the candidate solution of the above algorithm. Property (1) in definition~\ref{USJE} is easy to verify. Due to finite precision arithmetics, this property will not hold exactly, but the error is typically very small, that is, below $ 10^{-10} $ in the numerical examples that follow in Section~\ref{secolg}. Property (2) in the definition is a condition on an infinite number of points and thus cannot be verified numerically. We simulate the economy for a very large number of periods (in the examples below for $ m=100\cdot n $ periods) and check the condition along the simulated path. If it is violated, one needs to run Algorithm \ref{alg3} with larger $m$ or possibly larger $ \eta $ and a better method to find points that imply a large error.

\section{An OLG economy}
\label{secolg}

To illustrate our general results derived in Section~\ref{sec:model} in a specific application, we consider a pure exchange economy with overlapping generations of agents. We show that in this setting, Assumption \ref{ass2} follows from standard assumptions on economic fundamentals. Furthermore, we also illustrate our computational method with a concrete numerical example.

At each date event, a continuum of identical agents enter the economy and live for $A$  periods. We denote the set of all date events at time $t$ by $ {\mathbf Z}^t $ and, taking $ z_0 $ as fixed, we write $ z^t \in {\mathbf Z}^t $ for any $ t \in \Nat_0 $ (including $ t=0 $). At each $ z^t$, there are finitely many different agents actively trading (distinguishing themselves by the age and the history of shocks), who are collected in a set $ {\mathbf I}=\cup_{a=1}^A  $. A specific agent at a given node $z^t$ is denoted by $ a \in {\mathbf I}$.

At each date event, there is a single perishable commodity,
the individual endowments are denoted by $ e_{a}(z^{t}) \in \Rea_{+} $, and are assumed to be time-invariant and measurable functions of the current aggregate shock.
Each agent who can be identified by his date event of birth, $ z^t $, has a time-separable expected utility function, that is,
\begin{equation*}
	U_{z^t}\left((c_{t+a})_{a=0}^{A-1}\right) = \Ex{t}{ \sum_{a=1}^{A}  u_{a}\left( c_{z^t,t+a-1} \right)},
\end{equation*}
where  $c_{z^t,t+a-1} \in \Rea_+ $ denotes the agent's (stochastic) consumption at date $t+a-1$. 

There are $J$ assets, $ j \in {\mathbf J}=\{1,\ldots,J\} $, traded at each date event. The assets can be infinitely lived Lucas trees in unit net supply, or one-period financial assets in zero net supply. The net supply of an asset $j$ is denoted by $ \bar \theta_j \in \{0,1 \} $. 
Each asset $j$ is traded at a price $q_j $, and its (non-negative) payoffs depends on the aggregate shock and possibly on the current prices of the assets, that is, $  f_j: \Rea^J_+ \times {\mathbf Z}  \rightarrow \Rea_+ $. If asset $j$ is a Lucas tree (i.e., an asset in positive net supply), then
$ f_j(q,z)=q_j+d_j(z) $ for some dividends $ d_j: {\mathbf Z} \rightarrow \Rea_{+} $. Asset $j$ could also be a collateralized loan whose payoff depends on the value of the underlying collateral, or an option, or simply a risk-free asset. The aggregate dividends of the trees are defined as $ d(z)=\bar \theta \cdot f(q,z) - \bar \theta \cdot q $. 
The aggregate consumption is then $ e(z)=d(z) + \sum_{a=1}^A e_a(z) $. 
An agent $a$ faces trading constraints $ \theta \in {\mathbf \Theta}_{a} \subset \Rea^J $, where $ {\mathbf \Theta}_{A} = \{0\} $.
To simplify the notation, we write $ \vec \theta = (\theta_{a})_{a \in {\mathbf I}} $, $ \vec \theta^- = (\theta^-_{a})_{a \in {\mathbf I}} $ and
$ \vec c= (c_{a})_{a \in {\mathbf I}} $.

It is useful to define the set of possible portfolio holdings with market-clearing built-in as follows:
$$ {\mathbf \Theta}=\{\vec \theta :
\sum_{a \in {\mathbf I}}  \theta_{a}=\bar \theta, \quad \theta_{a} \in {\mathbf \Theta}_{a} \mbox{ for all } a-1 \in {\mathbf I} \} . $$
Similarly let the set of all beginning-of-period portfolio holdings be
$$ {\mathbf \Theta}^-=\{ \vec \theta^- : \theta^-_{1}=0, 
\sum_{a-1 \in {\mathbf I}}  \theta^-_{a}=\bar \theta \mbox{ and } \theta^-_{a} \in {\mathbf \Theta}_{a-1} \mbox{ for all } a \}.  $$ 

\subsection{Uniformly self-justified equilibrium}
\label{sec:SJE}

It is straightforward to embed this OLG model into our abstract framework.
The endogenous state is $ x=\vec \theta^- \in {\mathbf \Theta^-} = {\mathbf X} $, current endogenous variables are given by $ y=(q, \vec c, \vec \theta) $. The expectations correspondence can be defined as follows:
$$ (x_1,y_1\ldots x_Z,y_Z) \in {\mathbf H(\hat x, \hat y , \hat z)} \mbox{ if and only if } $$\
	each  $ (c_a,\theta_a) $,  $ a=1,\ldots,A$  solves the following problem:
	\begin{eqnarray}
	\label{ecolg}
		&& \max_{\theta \in {\mathbf \Theta_a},c\ge 0} u_a(c)+ \beta \sum_{z' \in {\mathbf Z}} \pi(z'|z) u_{a+1}'(e_{a+1}(z')+ \theta \cdot f(q_{z'},z') - q_{z'} \theta_{z'}) f(q_{z'},z') \cdot \theta \\ \notag
		&&  \mbox{s.t. } c=e_a(z)+\theta^-_a \cdot f(q,z) -q \cdot \theta  \quad.
	\end{eqnarray}

Since agents solve finite-dimensional convex optimization problems, the first-order conditions are necessary and sufficient, and the expectations correspondence characterizes competitive equilibria.

The transition function is given by $ T_{z'}(x,y,z)=(\theta_{1},\ldots \theta_{A-1}) $, and a policy function maps $ {\mathbf \Theta}^- $ to current prices, consumption-levels, and new asset holdings across agents.
A self-justified equilibrium is defined as above (definition~\ref{USJE}). The agents' expectations about the next period's asset prices and optimal choices are not necessarily correct, but the functions chosen are the best uniform approximation to actual equilibrium values.
For the definition, agents do not need to forecast portfolio choices in the subsequent period. It suffices to have forecasts for asset payoffs $ f(q,z) $ as well as for individual consumptions, $c$. Since we impose an Inada condition on utility, we want
to ensure that forecasted consumption is bounded away from zero. Therefore, our definition of the equilibrium correspondence in the context of the OLG model becomes

\begin{eqnarray*} && {\mathbf N}_P(\vec \theta^-,z) = \{ (q,\vec c, \vec \theta) : 
	\mbox{For all }a=1,\ldots,A-1 \\
 	&&  
(c_a,\theta_a) \in \arg\max_{\theta \in {\mathbf \Theta_a},c\ge 0} u_a(c)+ \beta \sum_{z' \in {\mathbf Z}} \pi(z'|z) u_{a+1}'(\hat P_{c_{a+1}}(T(\vec \theta),z')) f(P_q(T(\vec \theta),z'),z') \cdot \theta \\
&&  \mbox{s.t. } c=e_a(z)+\theta^-_a \cdot f(q,z) -q \cdot \theta \},
\end{eqnarray*}
where 
\begin{equation}
	\label{bish}
	 \hat{P}_{c_a}(\vec \theta^-,z) = \left\{ \begin{array}{ll} 
	\epsilon & \mbox{ if }  P_{c_a}(\vec \theta^-,z)\le \epsilon\\
P_{c_a}(\vec \theta^-,z)  & \mbox{ if } e(z)> P_{c_a}(\vec \theta^-,z) > \epsilon\\
e(z) & \mbox{ if }  P_{c_a}(\vec \theta^-,z)\ge e(z).
\end{array} \right. \end{equation}

In this setting, agents forecast the next period's optimal consumption and prices. This turns out to be computationally advantageous since consumption functions often do not exhibit strong nonlinearities.
Alternatively, we could assume that agents forecast marginal utilities.

Next, we show that the general existence proof for Theorem~\ref{thm1} readily gives sufficient conditions for existence in this model. Moreover, we illustrate our algorithm in the context of a simple numerical example.

\subsection{Existence in OLG exchange economies}
\label{Sec:main_result1}
In order to establish existence of a USJE, we need assumptions on fundamentals that guarantee the high-level Assumption~\ref{ass2} above holds:

 \begin{assumption} \mbox{}
 \label{ass3}
	\begin{enumerate}
		\item For each $a\in {\mathbf I} $, the Bernoulli-utility function $ u_{a}(\cdot) $ is continuously differentiable, strictly increasing, strictly concave, and satisfies an Inada condition, 
		$$ u_{a}'(x) \rightarrow \infty \mbox{ as } x \rightarrow 0 ,$$ and individual 
		endowments are positive, that is,
		$$ e_{a}(z) > 0 \mbox{ for all }  z \in {\mathbf Z} .$$
		\item The set $ {\mathbf \Theta} $ is compact, and for
		each $ a \in {\mathbf I} $, the set $ {\mathbf \Theta}_{a} $ is a closed convex cone  containing $ \Rea^J_+ $.
		\item The payoff functions, $ f: \Rea^J_+ \times {\mathbf Z} \rightarrow \Rea^J $, are non-negative valued and continuous.
		Moreover, for any $ i,j=1,\ldots,J $, the payoff $ f_j(q) $ only depends on $ q_i $ if $ \bar \theta_i > 0 $.
		\item For all $ \theta^- \in \Theta_{a-1} $,  
		$$ \theta^-_{a} \cdot f(q,z) \ge 0 \mbox{ for all } q \in \Rea^J_+, z \in {\mathbf Z} .$$
		\item For all $ k=1,\ldots, Y $, the basis functions, $ \psi_{k1},\ldots,\psi_{kD} $, are linearly independent.
	\end{enumerate}
\end{assumption}
Assumptions 3.1-3.3 are standard. Assumption 3.4 is motivated by collateral and default. Constraints ensure that agents cannot borrow against future endowments. In our formulation, this is true independently of prices---we implicitly allow for default (see, e.g.,~\cite{kubler2003stationary}).  

With these assumptions, the existence of a uniformly self-justified equilibrium simply reduces to the existence of a finite-dimensional fixed point, in particular, the main result of this section is as follows:
\begin{lemma}
\label{lemfi}
Given any finite set $ \widehat{\mathbf \Theta}^- \subset {\mathbf \Theta}^- $, there exist $ P_k \in {\mathbf P}_k $, $ k=1,\ldots,Y$, and $ y( \theta^-,z) $ for all $ z \in {\mathbf Z}$, $ \theta^- \in\widehat{\mathbf \Theta}^-  $ that solve Equations~\eqref{ecolg} such that
$$ P_k \in \arg\min_{f \in {\mathbf P}_k} \sup_{\theta^- \in \widehat{\mathbf \Theta}^- } \| y(\theta^-,z)-f(x) \| .$$

\end{lemma}
{\bf Proof:}

Suppose that $ \widehat{\mathbf \Theta}^- $ contains $G$ elements, and let $ \widehat{\mathbf S} = {\mathbf Z} \times \widehat{\mathbf \Theta}^-$
and take prices at each $ s \in \widehat {\mathbf S} $ to lie in the trimmed simplex $ (p,q) \in \Delta_{\epsilon}^J=\{(p,q) \in \Rea^{J+1}_+, p + \sum_{j=1}^J q_j=1, p \ge \epsilon, q_j \ge \epsilon, j=1,\ldots, J \} $. 
We decompose the economy into sub-economies for each $ s \in \widehat{\mathbf S} $ and construct a map from a compact and convex set of all agents' choices, prices, and forecasts, $P $, into itself.
We show that this map is upper hemi-continuous and convex valued, and using Kakutani's theorem (see~\cite{border1985fixed}), we can show that this map has a fixed point. As $ \epsilon $ becomes sufficiently small, one can prove market-clearing. 

First, we need to find a suitable, convex, and compact domain for the map. Assumption~\ref{ass1} guarantees the existence of a compact set
$ {\mathbf T} \subset \Rea_+ \times \Rea^J $, 
with $$ \{ (c , \theta) \in \Rea_+ \times {\mathbf \Theta}_{h} :
(c - e_{h}(z))+\theta \cdot \frac{1}{p}q    -  \theta^-_{h} \cdot f(\frac{1}{p}q,z) \le 0 \} \subset {\mathbf T}, $$ for all
$ (p,q) \in \Delta^J_{\epsilon} $ and all $ h \in {\mathbf H}.$

We construct an upper hemi-continuous, non-empty and convex-valued correspondence, $ {\mathbf \Phi}$, mapping choices, prices and forecasts at each element in $ \widehat {\mathbf S} $  
to itself, which has a fixed point, that is,
\begin{equation}
	{\mathbf \Phi}: {\mathbf T}^{ZG (A-1)}    \times (\Delta^{J}_{\epsilon})^{ZG}  \times
	\Rea^{ZG(A-1)}_+
	\rightrightarrows  {\mathbf T}^{ZG (A-1) } \times (\Delta^{J}_{\epsilon})^{ZG} \times \Rea^{ZG(A-1)} .
\end{equation}

For this construction, 
for all $ a \in {\mathbf I} $ and all $s=(z,\vec \theta^-) \in \widehat {\mathbf S}$, agents take prices and forecasts as given, and their best response correspondence is given by
\begin{eqnarray} && {\mathbf \Phi}_{h,s}((x_t,p_t,q_t)_{t \in \widehat{\mathbf S}}) = \\ && \arg\max _{(c, \theta) \in {\mathbf T}} u_{h}(x)+\beta_h \sum_{z' \in {\mathbf Z}} \pi(z'|z)   u_{a+1}'(\hat P_{c_{a+1}}(T(\vec \theta),z')) f(P_q(T(\vec \theta),z'),z')\cdot \theta  \\ & & \quad\quad   \mbox{ s.t. } \nonumber \\
	&&  \quad  \quad (c - e_{h}(z))+\theta \cdot \frac{1}{p_s}q_s    -  \theta^-_{h} \cdot f(\frac{1}{p_s}q_s,z) \le 0,\nonumber
\end{eqnarray}
with $ \hat P $ being defined as above in Equation~\eqref{bish}.
The forecasts are formed by an \lq\lq artificial\rq\rq \  agent  who takes choices and prices
$ y(z,\vec \theta) $, $ z \in {\mathbf Z}$, $ \vec \theta \in \widetilde{\mathbf \Theta} $ as given, and solves for all $k=1,\ldots, Y$,
\begin{equation}
	\label{fpforc}
	P_k(z,\cdot)=
	\arg\min_{f \in {\mathbf P}}   \sup_{\vec \theta \in \widehat{\mathbf \Theta}} | P(\vec  \theta) -  y(z,\vec \theta) |.
\end{equation}
By Assumption~\ref{ass3}, we can, without loss of generality, take for each agent $a \in {\mathbf I} $, all $ z\in {\mathbf Z} $ and all $ j \in {\mathbf J} $ $ \alpha \in  {\mathbf A} $, where $ {\mathbf A}$ is a compact and convex set. Therefore, the best uniform approximation always exists, that is, equation~\eqref{fpforc} always has a solution.
Moreover, since $ {\mathbf A} $ is convex, and we minimize a convex function, the set of solutions is convex. Note also that by the maximum principle, the set of solutions is upper hemi-continuous in the $ZG $ points $ (c,q)(s) $, $ s \in \widehat{\mathbf S} $.
Therefore the map of forecasts,
\begin{equation}
	{\mathbf \Phi}_{H+1} : \Rea^{ZGHJ} \rightrightarrows \Rea^{ZGHJ},
\end{equation}
is upper hemi-continuous, convex valued, and non-empty valued.

For each $ s \in \widehat{\mathbf S} $, define the price-player's best response as follows:
\begin{equation}
	{\mathbf \Phi}_{0,s}(\vec \theta_{s},\vec x_{s})=\arg\max_{(p,q)\in {\mathbf \Delta}_{\epsilon}^J} p (\sum_{a \in {\mathbf I}} ( c_{a,s}-e_{a}(z)-d(z)))  + q \cdot (\sum_{a \in {\mathbf I}} ( \theta _{a,s}-\bar \theta )). 
\end{equation}
Assumptions \ref{ass1} and \ref{ass3} therefore, guarantee that the mapping 
\begin{equation}
	{\mathbf \Phi} = \times_{s \in {\mathbf S}, a \in {\mathbf I}}  {\mathbf \Phi}_{a,s} 
	\times_{s \in {\mathbf S}} 
	{\mathbf \Phi}_{0,s}    \times  {\mathbf \Phi}_{H+1}
\end{equation}
is non-empty, convex valued, and upper hemi-continuous. By Kakutani's fixed point theorem, there exists a fixed point with prices $ (\bar p_s, \bar q_s)_{s \in \widehat{\mathbf S}} $.

As $ \epsilon \rightarrow 0 $, Assumption \ref{ass3} guarantees that there will be a strictly positive $ \epsilon $ such that for all $ s \in \widehat{\mathbf S}$, that is,
\begin{equation}
	(\bar p_s, \bar q_s) \in 
	\arg\max_{(p,q)\in {\mathbf \Delta}_{0}^J} p (\sum_{a \in {\mathbf I}} ( c_{a,s}-e_{a}(z)-d(z)))  + q \cdot (\sum_{a \in {\mathbf I}} ( \theta _{a,s}-\bar \theta )) .
\end{equation}
By a standard argument, markets clear.
This finishes our proof. $\Box $\\

Note that above's proof implies that there exist convex and compact sets $ {\mathbf A} $ for each $ h \in {\mathbf H} $ such that for any finite equilibrium and any discretization $  \widehat{\mathbf \Theta} $, all forecasting coefficients for agent $h$ lie in this set.
This follows from the assumption that $ {\mathbf \Theta} $ is compact.

As a final lemma, we need to establish the following closed graph characterization of the equilibrium correspondence. The result follows directly from the definition of the temporary equilibrium.
\begin{lemma}
	\label{contin}
	Given a sequence of policy functions characterized by coefficients 
	$ \vec \alpha $ 
	converging to some $ \vec \alpha^* $ as $ i \rightarrow \infty $, let
	$  P^i $ and $ P^* $ denote the associated policy functions.
	For any $ z \in {\mathbf Z} $, consider  any sequence $ (\vec \theta_i^-, y_i) $ with 
	$ \vec \theta_i^- \in {\mathbf \Theta}^- $
	and $ y_i \in {\mathbf N}_{P^i} (z,\vec \theta_i^-) $ for all $i$, converging to
	$ (\vec \theta^{-*}, y^*) $.
	Then, the following holds: 
	$$ y^*  \in {\mathbf N}_{P^*}(z,\vec \theta^{*-}) . $$
\end{lemma}

Now, our existence theorem (cf. theorem~\ref{thm1}) can be applied to ensure the existence of a USJE for the OLG economy.
Note that the key result, Lemma~\ref{lemfi}, can be proven with standard methods and that similar results can be established in models with production, and models with infinitely lived agents.

\subsection{Computation in an OLG economy: numerical examples}

To illustrate our numerical method proposed in Section~\ref{sec:comp-USJE}, we consider a simple example with $ A=10 $, and a single risk-free asset being traded each period. Agents face short-sale constraints, that is,
$$ \theta_a \ge - \min_{z \in {\mathbf Z}} e_{a+1}(z). $$

The purpose of the example is not to claim that the computational method is competitive compared to other existing methods in terms of performance and speed, but rather to illustrate that i) USJE can be computed in non-trivial models and ii) to show that our concept of a USJE allows for straightforward error analysis of the candidate solution. While the example is almost trivial with respect to the asset structure, the endogenous state space is 8-dimensional\footnote{The old (a=10) do not trade, and in a pure exchange setting, asset holdings of generation $A-1$ follow from market clearing.} which makes it non-trivial to solve for standard collocation methods (see, e.g.,~\cite{Krueger20041411,SG_in_econ_handbook}, and references therein). We choose the stochastic process for individual endowments to ensure that the model exhibits complex equilibrium dynamics.

Individual endowments have a deterministic part, where endowments initially increase in age and then decrease,
$$ \bar e_{a}=\left\{ \begin{array}{ll}  
0.6+0.2 a & 4 \ge a \ge 1\\
1.4-0.2 \cdot (a-5) & 10 \ge a \ge 5 ,\\ \end{array} \right. 
$$
and they are subject to three additive random variables, $ \sigma_1, \sigma_2, \sigma_3 $, that are i.i.d., and take
values $0.1$ and $-0.1$ with probability $1/2$, such that
$$ e_a(z)=\bar e_a + f_{1a} \sigma_1(z) + f_{2a} \sigma_2(z) + f_{3a} \sigma_3(z). $$
The presence of three independent shocks implies that $ {\mathbf Z} $ has $ 2^3=8 $ elements.
The values for the coefficients $ f_{ja} $ are reported in Table~\ref{tab1}.
  \begin{table}[!t]
			\begin{center}
				\begin{tabular}{|c|c|c|c|c|c|c|c|c|c|c|}
					\hline 
            Age:  & 1 & 2 & 3 & 4 & 5 & 6 & 7 & 8 & 9 & 10 \\
            \hline
            $f_1$ & 1 &  $ \sqrt{2/3} $ & $ \sqrt{1/3} $ & 0 & 0 & 0 & 0 & 0 & 0 & 0 \\
            $f_2$ & 0 &  $ \sqrt{1/3} $ & $ \sqrt{2/3} $ & 1 & $\sqrt{2/3}$ & $ \sqrt{1/3}  $ & 0 & 0 & 0  & 0 \\
            $f_3$ & 0 & 0 & 0 & 0 & $ \sqrt{1/3} $ & $\sqrt{2/3}$ & 1 & 0.5 & 0.25 & 0  \\ \hline
\end{tabular}
\end{center}
	\caption{Coefficients on the three shocks for all ages.}
			\label{tab1}
\end{table}
Bernoulli utilities are assumed to exhibit constant relative risk aversion, that is,
$$ u_a(c)=\beta^{a-1} \frac{c^{1-\gamma}}{1-\gamma} ,$$
with $ \beta=0.75 $ and $ \gamma=3 $.

In this relatively simple setting, it is sufficient to forecast the next period's consumption policy since the prices and new portfolio holdings in the 
next period do not enter the equilibrium conditions.
We consider three different specifications for the set of admissible policies, ${\mathbf P}$. First, we consider a very restricted set of basis functions where we postulate that consumption only depends on the individual's beginning of period asset holding, that is, 
$$ {\mathbf P}^1_{c_a} = \{ 1, \theta^-_a, (\theta^-_a)^2, (\theta^-_a)^3 \}  .$$
This, of course, neglects the effects of the wealth distribution on asset prices and optimal choices.
As an alternative, we therefore consider
$$ {\mathbf P}^2_{c_a} = \{ 1, \theta^-_2, \ldots, \theta^-_{A-1}, (\theta^-_a)^2, (\theta^-_a)^3 \}  .$$
In this specification, optimal choices can depend non-linearly on an agent's own asset-holdings, but they also depend (linearly) on all other agents holdings. It does not allow for non-zero cross derivatives.
It may be useful to consider a third specification of $ {\mathbf P} $, which captures some interaction terms. We consider therefore the following expression:
$$ {\mathbf P}^3_{c_a} = \{ 1, \theta^-_2, \ldots, \theta^-_{A-1}, \theta^-_a \theta^-_2, \ldots, \theta^-_a \theta^-_{A-1}, (\theta^-_a)^3 \}  .$$

Given the specific economic model, we vary Algorithm~\ref{alg3} slightly. In step $0$ of the method, we solve the model along a simulated path to obtain initial guesses for policy and points.
As explained above, in Step 2.(b), we search for a point with locally large errors by performing one gradient ascent step with back-tracking line-search (see, e.g., \cite{nocedal2006numerical}).

We choose $n=200$, $\eta=0.1$  and $ m=50\cdot n $.  In each of the three specification of admissible forecasts, it took around 50 iterations until convergence.

Tables~\ref{tab2},~\ref{tab3}, and~\ref{tab4} report the maximum errors for our three specification of admissible forecasting functions (in units of percent). For $ a=9 $, the forecasting errors are zero because consumption is a linear function of asset holdings for $ a=10 $.
  \begin{table}[!tb]
			\begin{center}
				\begin{tabular}{|c|c|c|c|c|c|c|c|c|c|}
					\hline 
					Agent & State 1 & State 2 & State 3 & State 4 & State 5 & State 6 & State 7 & State 8 \\ \hline
agent 1 & 
 8.087 &
 7.944 &
 8.332 &
 7.790 &
 8.323 &
 7.967 &
 8.261 &
 8.017 \\ \hline
agent 2 &
 7.324 & 
 7.599 & 
 7.745 & 
 7.471 & 
 7.462 & 
 7.340 & 
 7.829 & 
 7.710 \\ \hline
agent 3 & 
 4.710 & 
 4.445 & 
 5.000 &
 4.2500 & 
 4.4975 & 
 4.1751 & 
 4.3967 & 
 4.3171 \\ \hline
agent 4 &
 2.6036 & 
 2.5321 & 
 2.5536 & 
 2.8338 &
 2.3538 & 
 2.3337 &
 2.5918 & 
 2.4610 \\ \hline
agent 5 &
 2.3683 &
 2.3610 &
 2.0465 &
 2.0528 & 
 1.9328 &
 1.9679 &
 1.6598 &
 1.4314 \\ \hline
 agent 6 &
 1.3623 & 
 1.6008 &
 0.7356 &
 0.8931 &
 0.9760 &
 1.1620 &
 1.3327 &
 1.2368 \\ \hline
agent 7 & 
 0.8842 & 
 1.0090 &
 1.2079 &
 1.3238 &
 1.0501 &
 1.1062 &
 1.4654 &
 1.6153 \\ \hline
 agent 8 & 
 0.9123 & 
 1.1001 &
 1.1999 &
 1.2297 &
 1.0811 &
 1.2104 &
 1.43666 &
 1.4053 \\ \hline

\end{tabular}
\end{center}
	\caption{Maximum error for ${\mathbf P}^1_{c_a}$ ($ \times 10^{-2} $).}
			\label{tab2}
\end{table}

  \begin{table}[!bt]
			\begin{center}
				\begin{tabular}{|c|c|c|c|c|c|c|c|c|c|}
					\hline 
					Agent & State 1 & State 2 & State 3 & State 4 & State 5 & State 6 & State 7 & State 8 \\ \hline
agent 1 & 
 0.2494 &
 0.2341 &
 0.2619 &
 0.3009 &
 0.2080 &
 0.2570 &
 0.2764 &
 0.3247\\ \hline
agent 2 &
 0.3068 &
 0.3372 &
 0.3579 &
 0.3673 &
 0.3402 &
 0.3732 &
 0.4091 &
 0.4599 \\ \hline
agent 3 &
 0.3169 &
 0.4021 &
 0.3921 &
 0.4640 &
 0.3646 &
 0.4170 &
 0.4378 &
 0.4677\\ \hline
agent 4  &
 0.3361 &
 0.3714 &
 0.3927 &
 0.4650 &
 0.3739 &
 0.4410 &
 0.4058 &
 0.4173 \\ \hline
agent 5 &
 0.3603 &
 0.4373 &
 0.4426 &
 0.4641 &
 0.4111 &
 0.4615 &
 0.4583 &
 0.5686 \\ \hline
agent 6 &
 0.5492 &
 0.6472 &
 0.5485 &
 0.7110 &
 0.5752 &
 0.6557 &
 0.7281 &
 0.7092 \\ \hline
agent 7 &
 0.5844 &
 0.6057 &
 0.6499 &
 0.6416 &
 0.6399 &
 0.6336 &
 0.6497 &
 0.7246 \\ \hline
agent 8 &
 0.5685 &
 0.6025 &
 0.5725 &
 0.6890 &
 0.5312 &
 0.6511 &
 0.6345 &
 0.6647 \\ \hline

\end{tabular}
\end{center}
	\caption{Maximum error for  ${\mathbf P}^2_{c_a}$ ($ \times 10^{-2} $).}
			\label{tab3}
\end{table}

\begin{table}[!bt]
			\begin{center}
				\begin{tabular}{|c|c|c|c|c|c|c|c|c|c|}
					\hline 
					Agent & State 1 & State 2 & State 3 & State 4 & State 5 & State 6 & State 7 & State 8 \\ \hline
agent 1 &
 0.0716 & 
 0.0906 & 
 0.1069 &
 0.0915 &
 0.1011 &
 0.1285 &
 0.1294 &
 0.1860\\ \hline
agent 2 &
 0.0655 & 
 0.0873 &
 0.0914 &
 0.1293 &
 0.0854 &
 0.1297 &
 0.1006 &
 0.1400\\ \hline
 agent 3 & 
 0.0643  &
 0.0776 & 
 0.0790 &
 0.0999 &
 0.0669 &
 0.0941 &
 0.0849 &
 0.1056 \\ \hline
agent 4 & 
 0.0706 &
 0.0693 & 
 0.0728 &
 0.0812 &
 0.0678 &
 0.0753 & 
 0.1033 &
 0.0863\\ \hline
agent 5 & 
 0.0976 & 
 0.1167 & 
 0.1260 &
 0.1415 & 
 0.1079 &
 0.1290 & 
 0.1186 &
 0.1330\\ \hline
agent 6 &
 0.1246 &
 0.1249 &
 0.1505 &
 0.1563 &
 0.1528 &
 0.1646 &
 0.1629 &
 0.1672 \\ \hline
agent 7 &
 0.1248 &
 0.1302 &
 0.1402 &
 0.1579 &
 0.1293 &
 0.1452 &
 0.1447 &
 0.1530\\ \hline
 agent 8 & 
  0.0461 &
  0.0384 &
  0.0537 &
  0.0448 &
  0.0382 & 
  0.0384 & 
  0.0409 &
  0.0420 \\ \hline
\end{tabular}
\end{center}
	\caption{Maximum error for  ${\mathbf P}^3_{c_a}$ ($ \times 10^{-2} $).}
			\label{tab4}
\end{table}

\begin{table}[!bt]
			\begin{center}
				\begin{tabular}{|c|c|c|c|c|c|c|c|c|}
					\hline 
					Agent 1 & Agent 2 & Agent 3 & Agent 4 & Agent 5 & Agent 6 & Agent 7 & Agent 8 \\ \hline
					0.0505 & 0.0406 & 0.0442 & 0.0419 & 0.0509 & 0.0767& 0.0597 & 0.0128\\ \hline
\end{tabular}
\end{center}
	\caption{Average error for ${\mathbf P}^3_{c_a}$ ($ \times 10^{-2} $).}
			\label{tab5}
\end{table}

As expected, adding more terms to the set of admissible functions increases the accuracy of the approximation. Interestingly the effects are quite different depending on the agent's age, $a$.
While for $ {\mathbf P}^1_{c_a} $ the errors are about 8 times higher for $ a=1 $ than $ a=8 $, for $ {\mathbf P}_{c_a}^2 $  it is a factor of about 2-3, and for $ {\mathbf P}_{c_a}^3 $ less than a factor of 3.

In Table \ref{tab5} we report the average error (along a simulated path) for the  set $ {\mathbf P}_{c_a}^3 $. As in the other two specifications, the average error is typically about 2-3 times lower than the maximal error. Note that since our approach minimizes the maximal error, the average error is relatively high compared to the maximal error (when compared to methods that minimize the average error).

An obvious shortcoming of our approach lies in the fact that the sets of admissible forecasts are exogenous. It would be useful to consider a loss function that chooses forecasts to trade off accuracy and the complexity of the forecast (see, e.g.,~\cite{kubler2019self}). It is worth noting that forecasting errors for $ {\mathbf P}^3_{c_a} $ are generally very small, and it seems difficult to argue in favor of the need for more accuracy. On the other hand, errors for the first specification are significant and imply substantial welfare losses for many agents. For both the examples considered in~\cite{Krueger20041411} and those in this paper, the wealth distribution has a substantial impact on prices, and non-trivial aspects of this wealth distribution need to be used for good forecasts.

\section{Conclusion}
\label{seccon}

We present a novel equilibrium concept for stochastic dynamic general equilibrium models with heterogeneous agents. A USJE always exists and can be approximated numerically with arbitrary precision.  The choice of the sup-norm plays an important technical role in our existence proof.  From an economic point of view, it is reasonable to assume that agents aim to minimize the maximum forecasting error. In particular, in a framework with overlapping generations, minimizing the average error (i.e., working with the computationally more convenient $ L^2 $ norm) allows for a situation where an agent makes large errors throughout his entire lifetime, hence experiencing large welfare losses).

Methods that use uniform function approximation, also referred to as Chebyshev approximation,\footnote{in honor of P. L. Chebyshev, who first studied it in \cite{chebychev1854orie}.}
min-max approximation, or $ L_{\infty}$ approximation, are no longer at the heart of computational science, mainly because they are often thought to be intractable.  However, as we show in this paper, for our purposes, it is not clear whether these methods compare favorably to least-squares regression.

Our primary motivations for deviating from rational expectations are i) that recursive rational expectations equilibria do not always exist, and ii) that a rigorous error analysis of a candidate numerical approximations to a rational expectations equilibrium is often not possible.
Our proposed alternative equilibrium, the USJE, always exists and allows for a simple numerical verification procedure.

The fact that USJE always exist and that they can be characterized by a finite system of non-linear equations and inequalities makes it amenable to formal investigations of local comparative statics and multiplicity. However, it is subject to further research to pursue this in more detail. 

 \bibliography{bib_econ}

\end{document}